\documentclass[3p,times,procedia]{elsarticle}
\usepackage{nupha_ecrc}

\volume{00}

\firstpage{1}

\journalname{Nuclear Physics A}

\runauth{}

\jid{nupha}

\jnltitlelogo{Nuclear Physics A}

\usepackage{amssymb}

\usepackage[figuresright]{rotating}

\begin{document}

\begin{frontmatter}



\dochead{XXVIIth International Conference on Ultrarelativistic Nucleus-Nucleus Collisions\\ (Quark Matter 2018)}

\title{Highlights from the LHCb experiment}


\author{Michael Winn on behalf of the LHCb collaboration}

\address{Laboratoire de l'acc\'{e}l\'{e}rateur lin\'{e}aire, Orsay}

\begin{abstract}
  We report recent results by the LHCb collaboration in heavy-ion collisions in collider and fixed-target mode at the LHC. 
  A large variety of measurements show the potential of LHCb in nuclear collisions.
  
\end{abstract}


\end{frontmatter}


\section{Introduction}
\label{sec:intro}
The LHCb experiment is designed for heavy-flavour measurements, precision tests of the standard model and searches for physics beyond the standard model. Its forward rapidity acceptance in $2<\eta<5$, its instrumentation with precision tracking and vertex determination, the charged hadron particle identification, the muon system and calorimetry combined with a flexible trigger system featuring a software level trigger with an input rate of about 1 MHz in $pp$ collisions make the experiment a versatile laboratory for studies aiming at a better understanding of strongly interacting matter in the laboratory~\cite{Aaij:2014jba}.

\begin{figure}
  \centering
  \includegraphics[width=.8\textwidth]{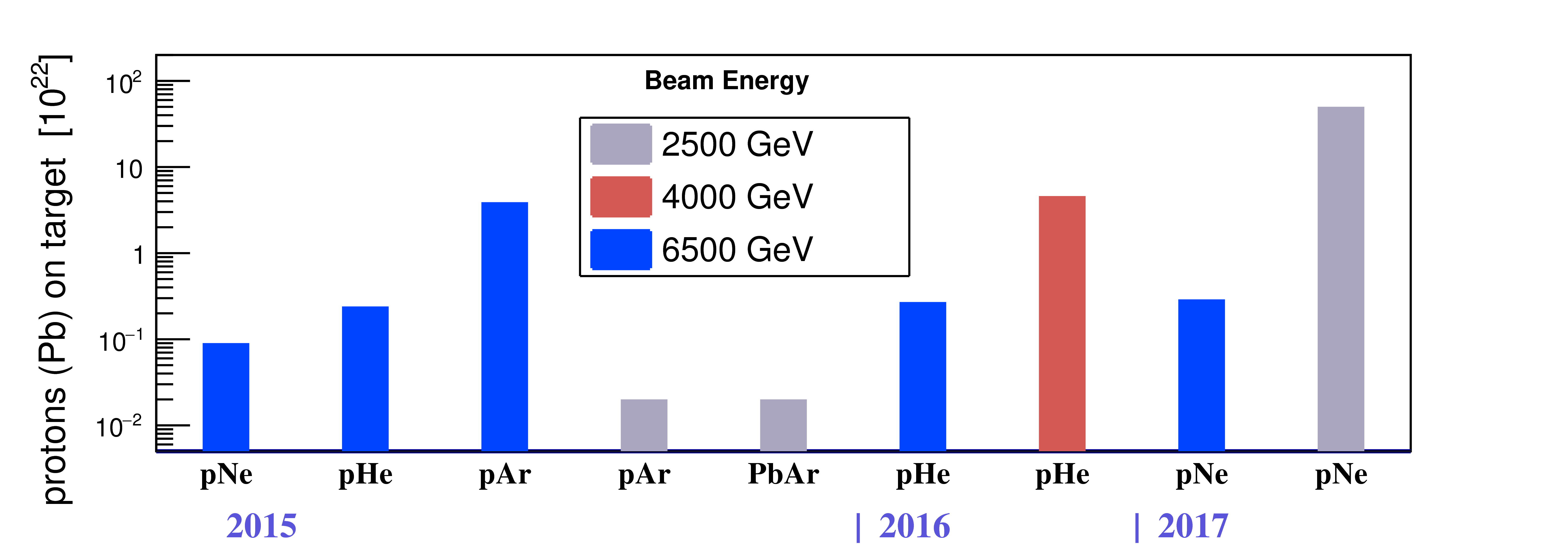}
  \caption{Available fixed-target data sets presented in terms of proton/ions on target.}
  \label{fig:fixedtarget}
\end{figure}

The LHCb collaboration is the youngest member of the LHC heavy-ion family participating with a small luminosity in the 2013 $p$Pb run at 5~TeV and in the 2015 PbPb run. Since these pioneering data takings, significantly larger data sets were successfully recorded and are being planned to be taken this year. Table~\ref{tab:coll} presents already taken and planned collider data and Figure~\ref{fig:fixedtarget} shows the so far available fixed target data. We summarise in the following results that are for the first time presented at a Quark Matter conference edition.

\begin{table}
  \centering
  \begin{tabular}{|c|c|c|c|l|}
    \hline
    & year & $\sqrt{s_{NN}}$ & &\\
       \hline
    $p$Pb/Pb$p$ & 2013 & 5.02 TeV & 1.6~nb$^{-1}$ &\\
    PbPb   & 2015 & 5.02 TeV & 10~$\mu$b$^{-1}$  &    \\
    $p$Pb$/$Pb$p$ & 2016 & 8.16 TeV & 34~nb$^{-1}$  &  $21$ $\times 2013$ \\
    XeXe   & 2017    & 5.44 TeV    & 0.4 $\mu$b$^{-1}$  &    \\
    \textit{PbPb}   & \textit{2018}& 5.02 TeV & goal: 100~$\mu$b$^{-1}$ & \textit{$10$ $\times 2015$} \\
    \hline
  \end{tabular}
      \caption{Overview of collision systems and recorded luminosities by LHCb in heavy-ion collisions in collider mode.}
      \label{tab:coll}
\end{table}

\section{Correlation measurements in $pp$ collisions}
\label{sec:corr}
\begin{figure}[h]
   \centering
  \includegraphics[width=0.4\textwidth]{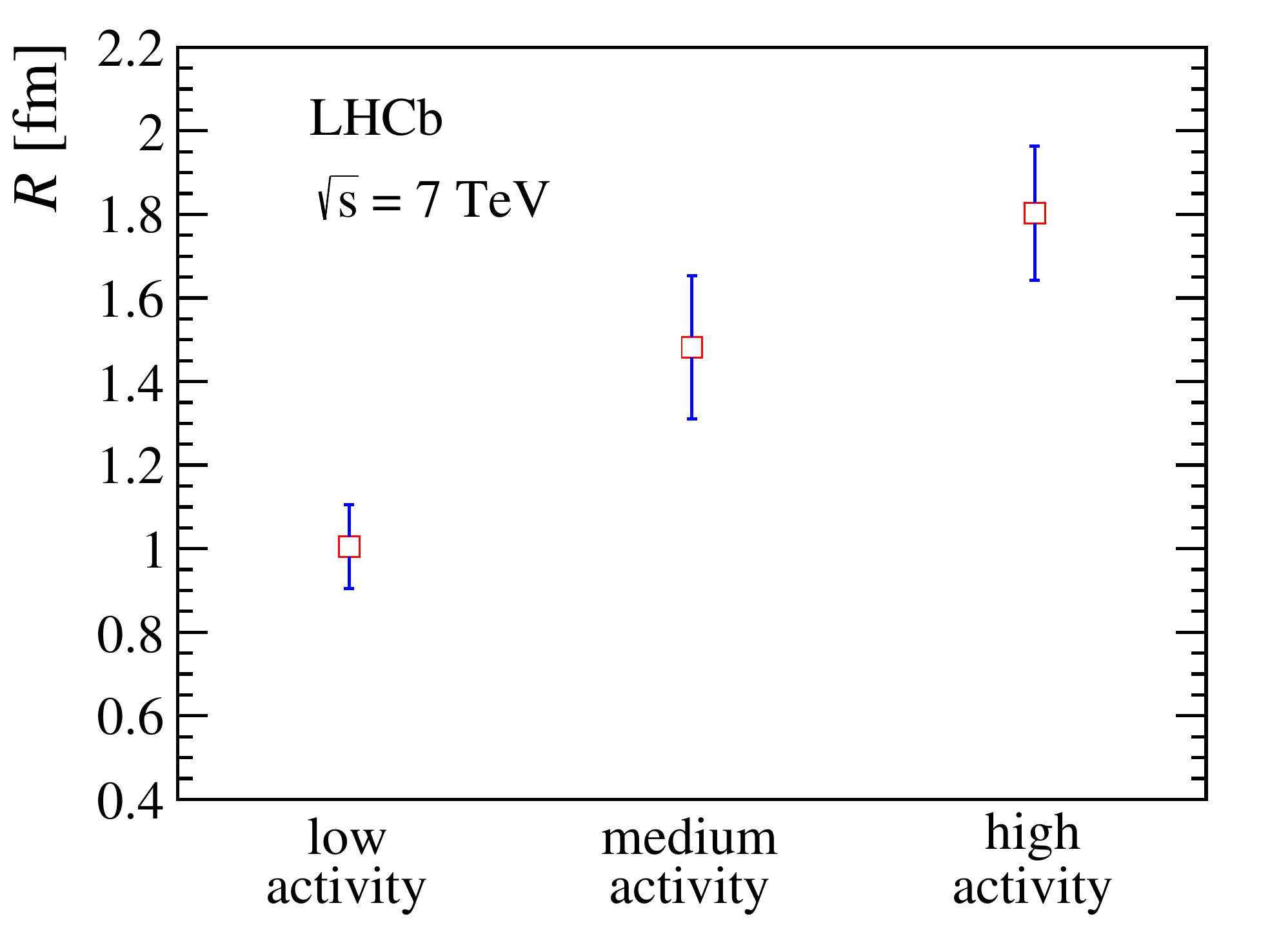}
  \includegraphics[width=0.4\textwidth]{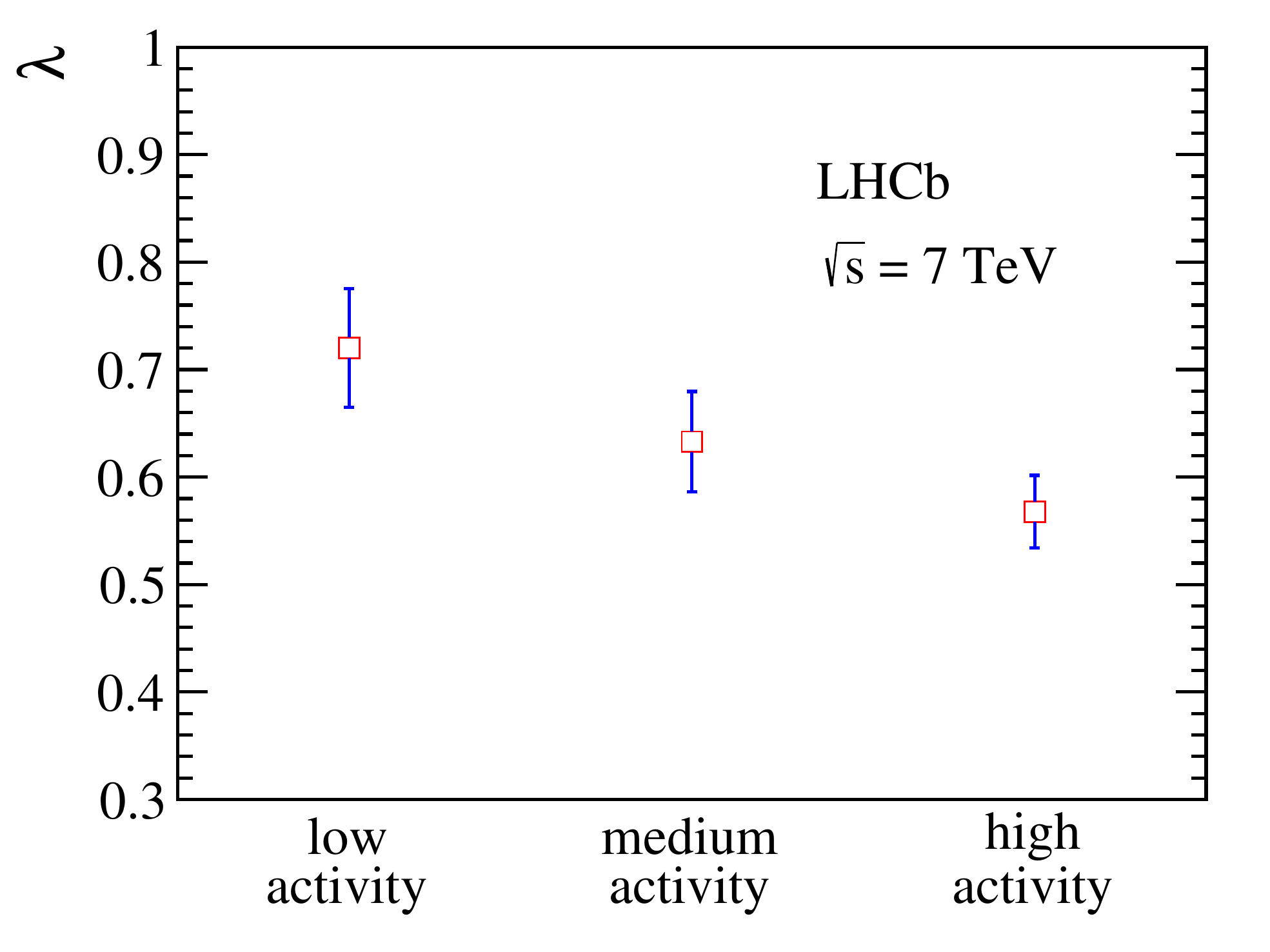}
  \caption{The system size parameter in fm and the chaoticity parameter as extracted in the three classes of event activity in the $pp$ analysis at $\sqrt{s}=$~7~TeV~\cite{HBT}. The activity classes correspond to charged particle multiplicity intervals $[8,18]$~(low), $[19,35]$~(medium), $[36,96]$~(high) of primary particles within the LHCb acceptance $2<\eta<5$ correcting the number of reconstructed tracks per primary vertex to the particle level with a Bayesian unfolding technique.  }
  \label{fig:HBT}
\end{figure}

LHCb proved its capability to perform dihadron correlation measurements of primary charged particles in $p$Pb collisions~\cite{corrpPb}. While this measurement type is also pursued in $pp$ collisions to probe collectivity~\cite{Kopecna:2017wee}, LHCb explored for the first time Bose-Einstein correlations of identical pions in $pp$ collisions to constrain the spatio-temporal particle emission patterns as a function of final state particle production multiplicity~\cite{HBT}. The correlation functions are presented in form of a double ratio of $C(Q) = \frac{C(Q)_2^{SE,data}/C(Q)_2^{ME,data}}{C(Q)_2^{SE,MC}/C(Q)_2^{ME,MC}}$ with $Q$ being the absolute value of the four-momentum vector difference between the two particles and the indices SE and ME indicating same event and mixed-event distributions. The correlation function is fitted with a function of the form $f(Q)=N\cdot(1+\lambda \cdot exp(- Q R))\cdot (1 + \delta Q)$, where $\lambda$ represent the so-called chaoticity and $R$ the source size.  Figure~\ref{fig:HBT} shows the best fit parameters.  The resulting source sizes show a similar dependence as observed at midrapidity by other LHC experiments~\cite{Khachatryan:2011hi, Aamodt:2011kd, Aad:2015sja} as a function of final state particle multiplicity.
 More differential studies and studies in $p$Pb collisions are natural extensions of this first result at forward rapidity at the LHC.

\section{Heavy-flavour production in $p$Pb collisions in collider mode}
\label{sec:HFpPb}
\begin{figure}
  \includegraphics[width=0.48\textwidth]{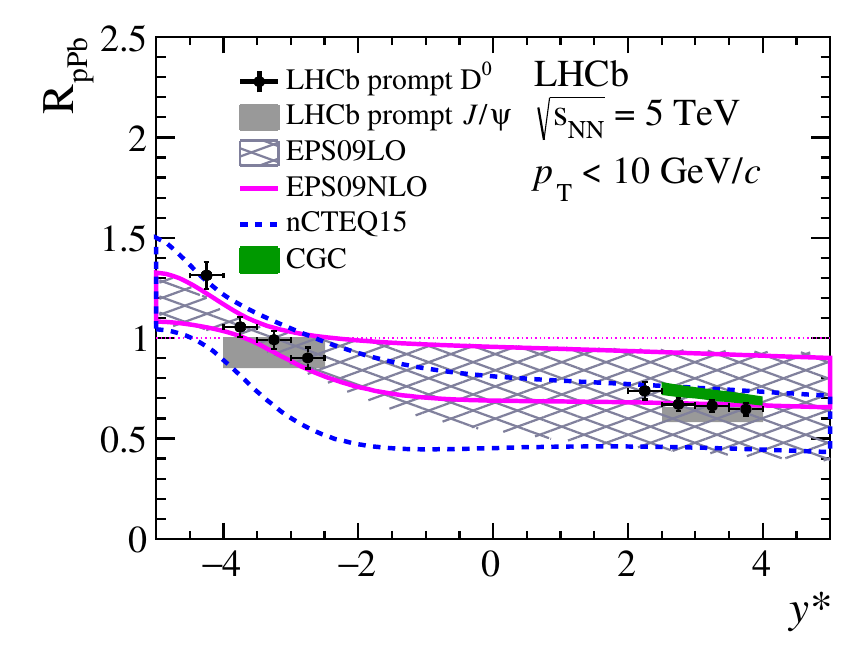}
  \includegraphics[width=0.48\textwidth]{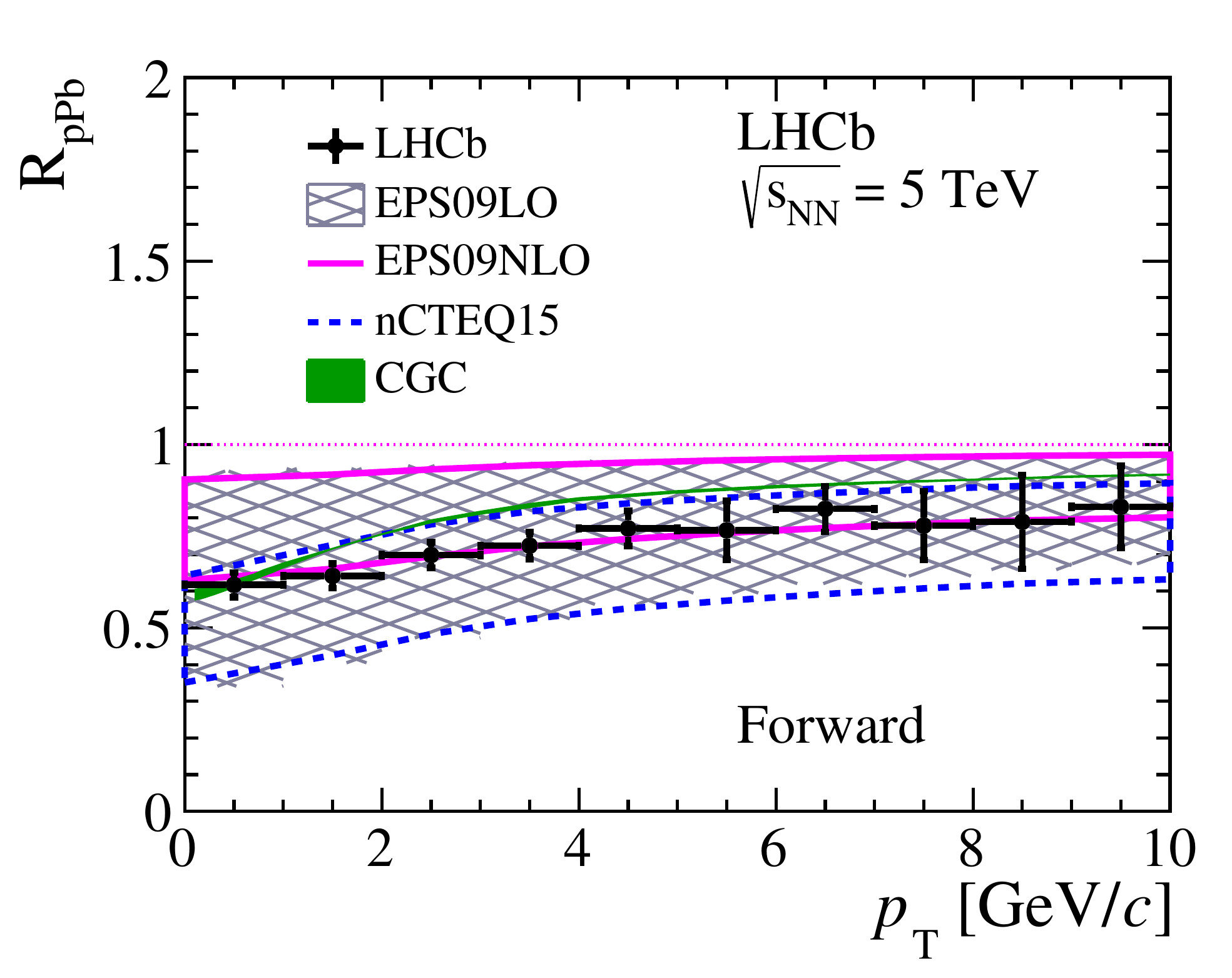}
  \caption{Left panel: the $D^0$ nuclear modification factor as a function of rapidity compared with model calculations in the collinear factorisation framework that are tuned to reproduce the $pp$ production cross sections and display the nPDF uncertainties on the left as well as a colour glass condensate calculation. Right hand: the nuclear modification factor at forward rapidity as a function of transverse momentum compared with the same model approaches detailed in Ref.~\cite{D0}.}
    \label{fig:D0}
  \end{figure}

The 2013 $p$Pb data set at $\sqrt{s_{NN}}=$~5~TeV allows for precise and unique production studies in the charm sector with LHCb. The observed prompt $D^0$ nuclear modification factor~\cite{D0} is shown in Figure~\ref{fig:D0}.  A suppression as low as 0.6 at low-$p_T$ at forward rapidity $2<y^*<4$ is observed. A slow approach to unity towards large transverse momenta of around 10~GeV/$c$ is visible. The data points are contained in the envelope of uncertainties within a model using collinear factorisation and being tuned to reproduce $pp$ collision data and displaying the uncertainties related to nuclear parton distribution functions (nPDF) that are considerably larger than the experimental uncertainties. The forward rapidity data, both integrated and $p_T$-differential, are reasonably reproduced by a  colour glass calculation in the dilute-dense approximation.  At backward rapidity, the last bins are indicating values above unity for rapidity differential nuclear modification factors.    Both observations, large modifications as well as the significantly smaller uncertainties than those affecting nPDF sets, urge and encourage us to clarify the sources of nuclear modifications in $p$Pb collisions. 
In particular, a clarification will be necessary to make full use of precision improvements of heavy-flavour observables in nucleus-nucleus collisions to extract parameters for quark matter studies. In addition, the characterisation of the gluonic content of nuclei and nucleons in view of physics of the saturation scale at the highest available collision energy would benefit from a falsification of available models via the combination of the full available set of data. 

\begin{figure}
    \includegraphics[width=0.49\textwidth]{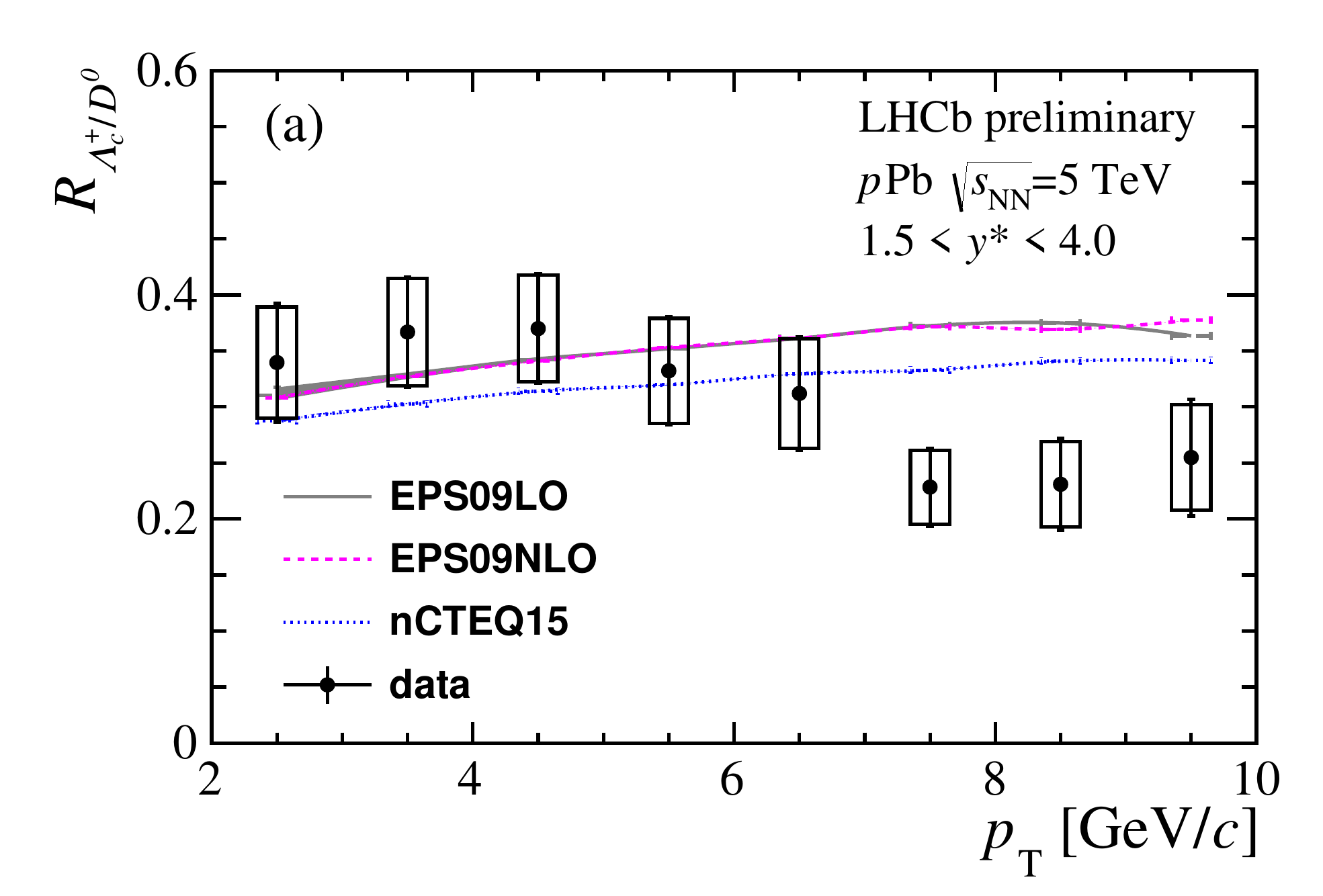}
    \includegraphics[width=0.49\textwidth]{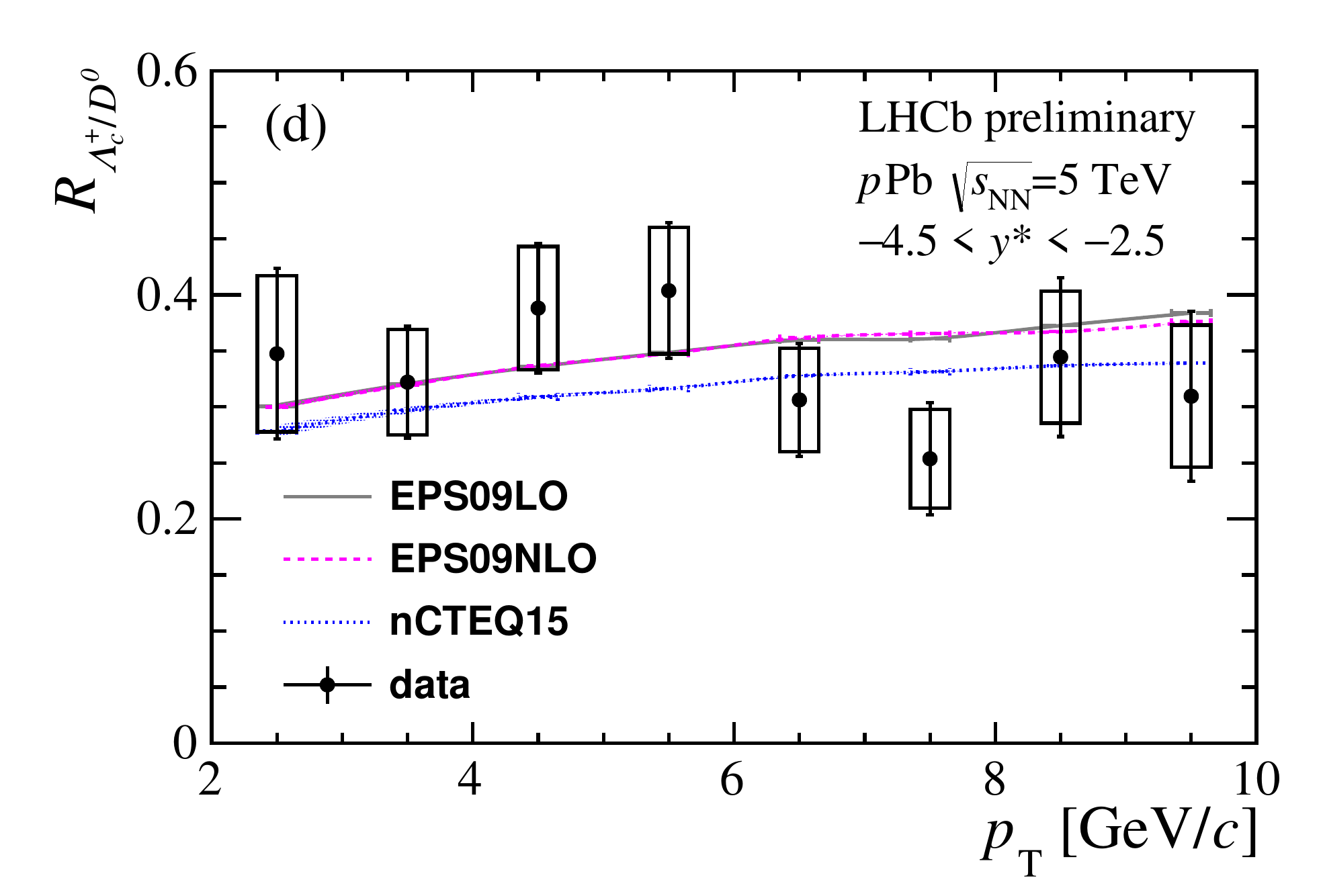}
    \caption{The prompt $\Lambda_C$ over $D_0$ production ratio at forward and at backward rapidity in $p$Pb collisions at $\sqrt{s_{NN}}=$~5 TeV measured by LHCb~\cite{Lc}.}
    \label{fig:Lc}
  \end{figure}

In $p$Pb collisions at $\sqrt{s_{NN}}= 5$~TeV, a first measurement of the prompt $\Lambda_C$ at forward and backward rapidity has been undertaken by LHCb\footnote{The results presented here will appear in~\cite{Lc}. They superseed the preliminary results published in 2017~\cite{LHCb:2017rvh}.}. In Figure~\ref{fig:Lc}, the prompt $\Lambda_C/D$ production ratio shows a behaviour consistent with a model calculation using $pp$ data as input for the parameterisation.  The measurement indicates similar values as observed recently by ALICE at midrapidity in $pp$ and in $p$Pb collisions~\cite{Acharya:2017kfy}.  The large observed values indicate tensions with existing fragmentation fraction measurements in $e^+e^-$ collisions and in $e$p collisions, see e.g. in ~\cite{Gladilin:2014tba} for a compilation of LEP data,  and warrant further tests of universality in charm hadronisation in $pp$ and $p$Pb collisions. Modifications of baryon-to-meson ratios have been observed in light flavour production in heavy-ion collisions compared to $pp$ collisions, see e.g. in~\cite{Abelev:2013xaa}. The observation of modification patterns in the heavy-flavour sector for baryon-to-meson ratios between different collision systems and their investigation as a function of charged-particle multiplicity in the final state may allow us to shed light into the modelling and understanding of these phenomena thanks to the presence of a mass scale much larger than $\Lambda_{QCD}$ for one of the valence quarks. In $p$Pb collisions, LHCb can contribute to this question at $\sqrt{s_{NN}}=$8.16~TeV with an integrated luminosity higher by a factor 20 with respect to the data at 5~TeV presented so far.

\begin{figure}
    \includegraphics[width=0.32\textwidth]{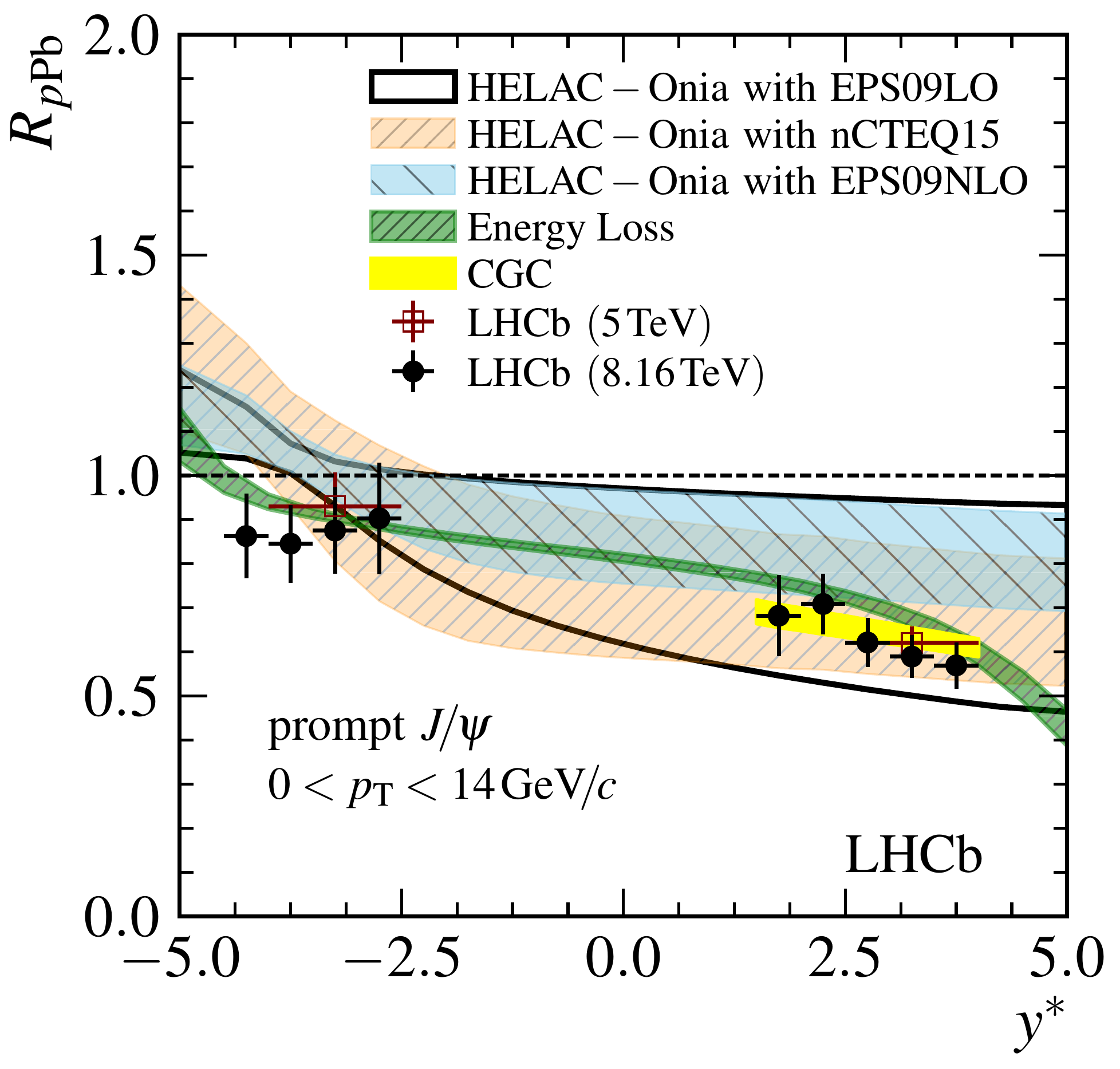}
    \includegraphics[width=0.32\textwidth]{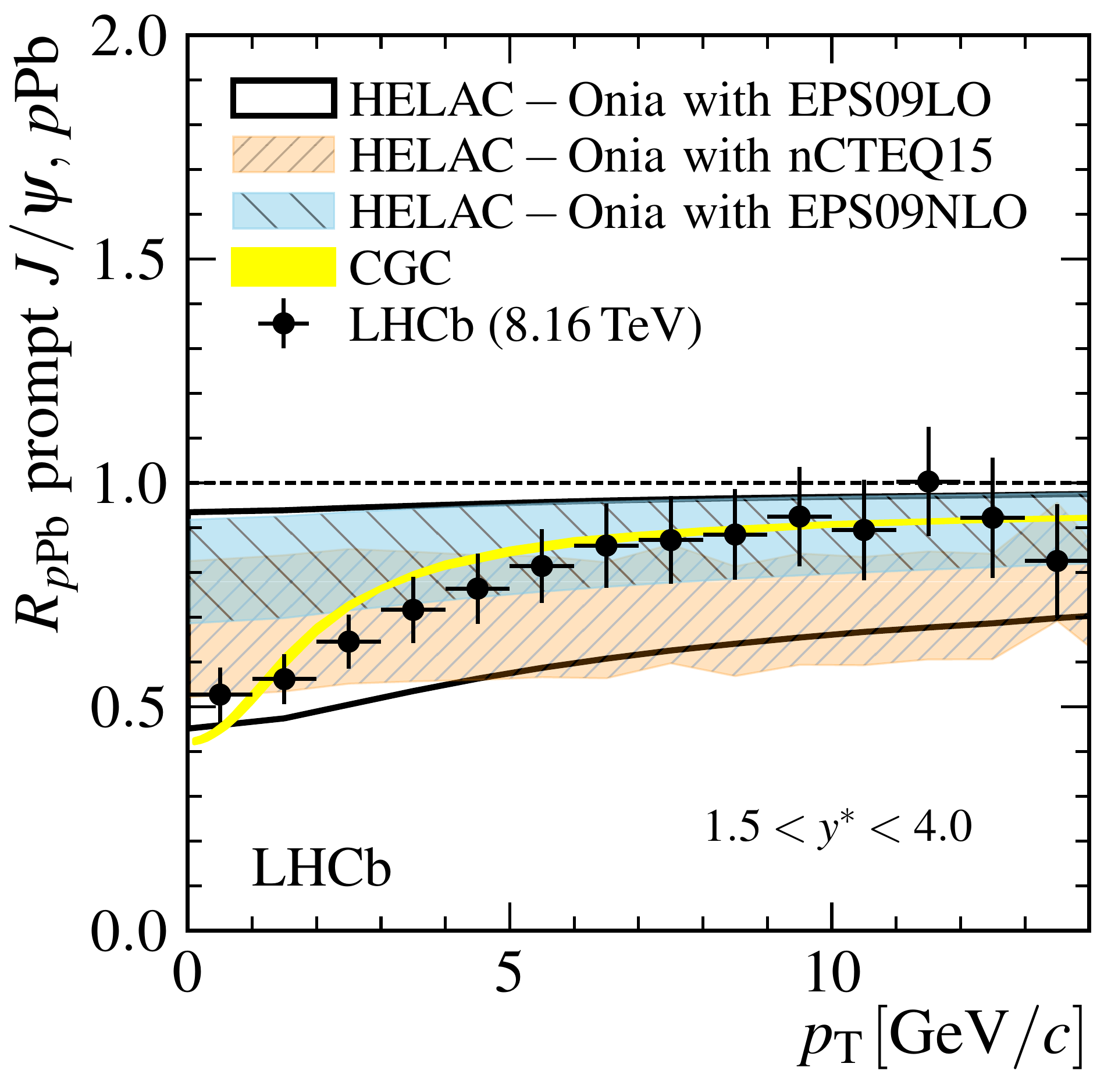}
    \includegraphics[width=0.32\textwidth]{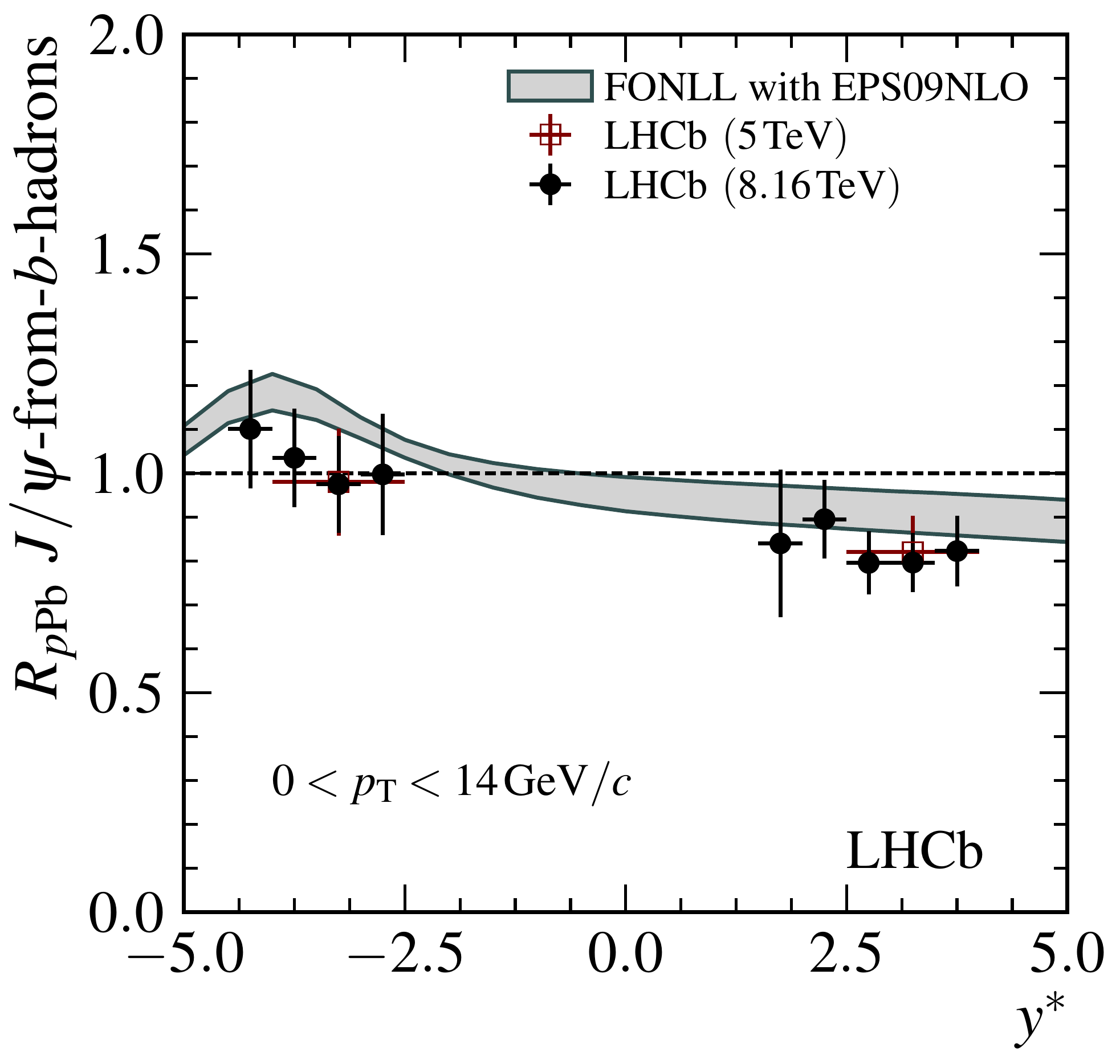}
    \caption{The prompt J/$\psi$ nuclear modification factor as a function of rapidity (left panel)  and of  transverse momentum (central panel) at $\sqrt{s_{NN}}=$8.16~TeV in $p$Pb collisions  at forward rapidity compared with model calculations. The nuclear modification factor of non-prompt J/$\psi$ as a function of rapidity integrated over transverse momentum~\cite{Jpsi} is shown in the right panel. Model references and detailed discussions are given in~\cite{Jpsi}. }
    \label{fig:Jpsi}
  \end{figure}

In the quarkonium sector, we report precise measurements of prompt and non-prompt J/$\psi$  in $p$Pb collisions at $\sqrt{s_{NN}}=$~8.16~TeV down to $p_T = 0$~\cite{Jpsi}. In Figure~\ref{fig:Jpsi}, the prompt J/$\psi$ production is compared with calculations using different nPDF sets, the coherent energy loss model as well as colour glass condensate calculations. Similar observations as for the open charm production apply: a strong suppression at low $p_T$ down to about 0.5 in the nuclear modification factor at forward rapidity is observed with an approach to unity at the highest transverse momenta and slight modifications much closer to unity at backward rapidity. There is no sign of an increase of the nuclear modification factor above unity as for the prompt $D^0$ at the backward edge of the acceptance. A possible physics origin of this apparent difference with respect to the $D^0$-result 
should wait for new measurements of the $D^0$ meson at $\sqrt{s_{NN}}=$~8.16~TeV. 

The non-prompt J/$\psi$ production shown in Figure~\ref{fig:Jpsi} (right panel) exhibits a weaker suppression at low $p_T$ in accordance with expectations for nPDF modifications. The non-prompt measurements provide the most precise test  of $b$-hadron production in nuclear collisions down to $p_T=0$ to date and are hence an important step to quantify modifications to be anticipated for the analysis of ion-ion data in the quest of modifications due to the presence of deconfined QCD matter in the open beauty and in the bottomonium sector.


  \section{Heavy-flavour production in fixed-target mode in $p$Ar and $p$He collisions}
  \label{sec:SMOG}

  The fixed target data sets of LHCb are a unique chance at the LHC to allow for precise measurements at centre-of-mass energies of around 100~GeV in the backward rapidity hemisphere in the nucleon-nucleon centre-of-mass system. We concentrate at this conference on charm production measurements in proton-induced collisions, that are interesting to constrain nuclear effects not related to deconfinement for upcoming analyses in nucleus-nucleus collisions as for example nPDFs as well as to constrain the nature of the partonic content of nuclei/nucleons. In particular, it has been argued that the proton contains higher excited Fock-state components with an additional valence-like $c\bar{c}$ quark pair~\cite{Brodsky:1980pb}. Due to the coverage getting close to the kinematic edge for heavy-flavour production at backward rapidity, the LHCb detector is well suited to have a closer look whether sizeable signs of this conjectured contribution are present.
  
  In 2017, LHCb reported for the first time measurements both in the soft sector~\cite{SMOGsoft} in view of cosmic ray physics as well as in the charm sector in $p$Ar collisions~\cite{LHCb:2017qap} in its unique fixed target configuration.

  At this conference, we present the results to be published in an upcoming article on D$^0$ and J/$\psi$ production in $p$He and $p$Ar collisions. The measurements in $p$Ar are hence superseeded by the results presented here. New results in $p$He featuring cross sections based on an indirect luminosity determination via elastic electron-proton scattering are shown.

\begin{figure}
    \centering
    \includegraphics[width=0.40\textwidth]{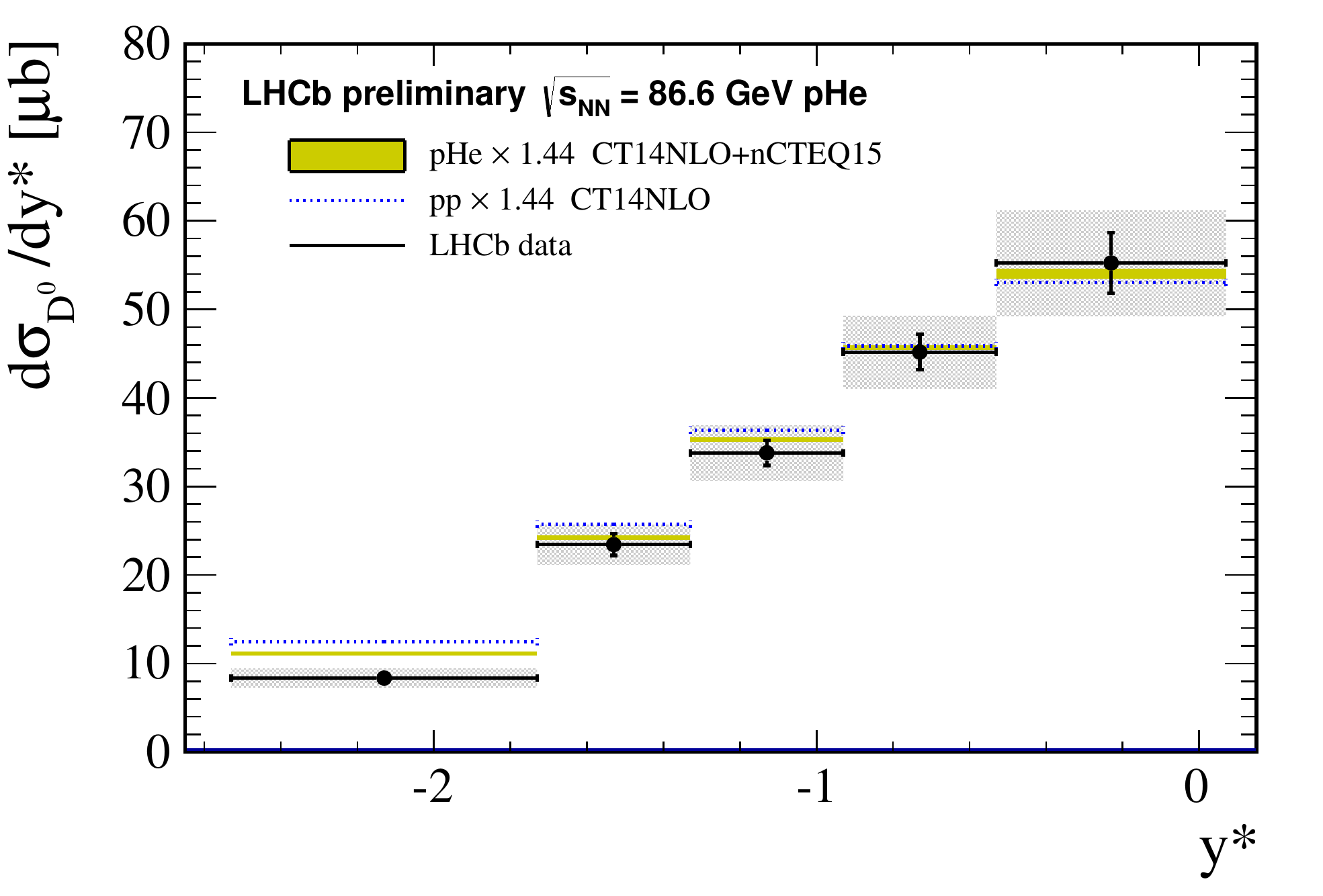}
    
    \caption{The $D^0$ production cross section as a function of rapidity compared with models, references detailed in~\cite{SMOGcharm}.}
    \label{fig:SMOG}
  \end{figure}

  As a  highlight from this data, Figure~\ref{fig:SMOG} shows the cross section of $D^0$ as a function of rapidity in $p$He collisions~\cite{SMOGcharm}. The shape is in reasonable agreement with model calculations not invoking an intrinsic charm contribution to the proton/nuclear wave function. We estimate the Bjorken-$x$ of the partons in the target nucleus. We use $x=\frac{2\cdot m_C}{\sqrt{s_{NN}}} \cdot e^{-y_{*}}$, where $m_C$, $\sqrt{s_{NN}}$, $y^{*}$ represent the charm quark mass, the centre-of-mass energy per nucleon-nucleon pair and the $D^0$ rapidity in the nucleon-nucleon centre-of-mass system. Hence, we conclude to be sensitive to a valence-like intrinsic charm contribution  in absence of strong nuclear effects for $^4$He\footnote{There could be conspirative cancellations of the intrinsic charm component and the EMC effect that is hardly constrained for nuclear gluon parton distribution functions by measurements.}. We see in the distribution rather a deficit than a relative enhancement of the production cross section at the backward rapidity edge with respect to the calculations without intrinsic charm. Therefore, we do not see  a valence-like intrinsic charm signal in our present data based on the distribution shape.

  \section{Towards direct photon measurements in $p$Pb collisions in collider mode}
  \label{sec:photon}

For suitable kinematics,  colour neutral final states can avoid to a large extent possible nuclear modifications that are not related to the partonic content of nuclei. In particular, direct photons are a suitable tool to probe the partonic structure at low-$x$ at forward rapidity at hadron colliders to search for signs of gluon saturation.
  
\begin{figure}
    \centering
    \includegraphics[width=0.40\textwidth]{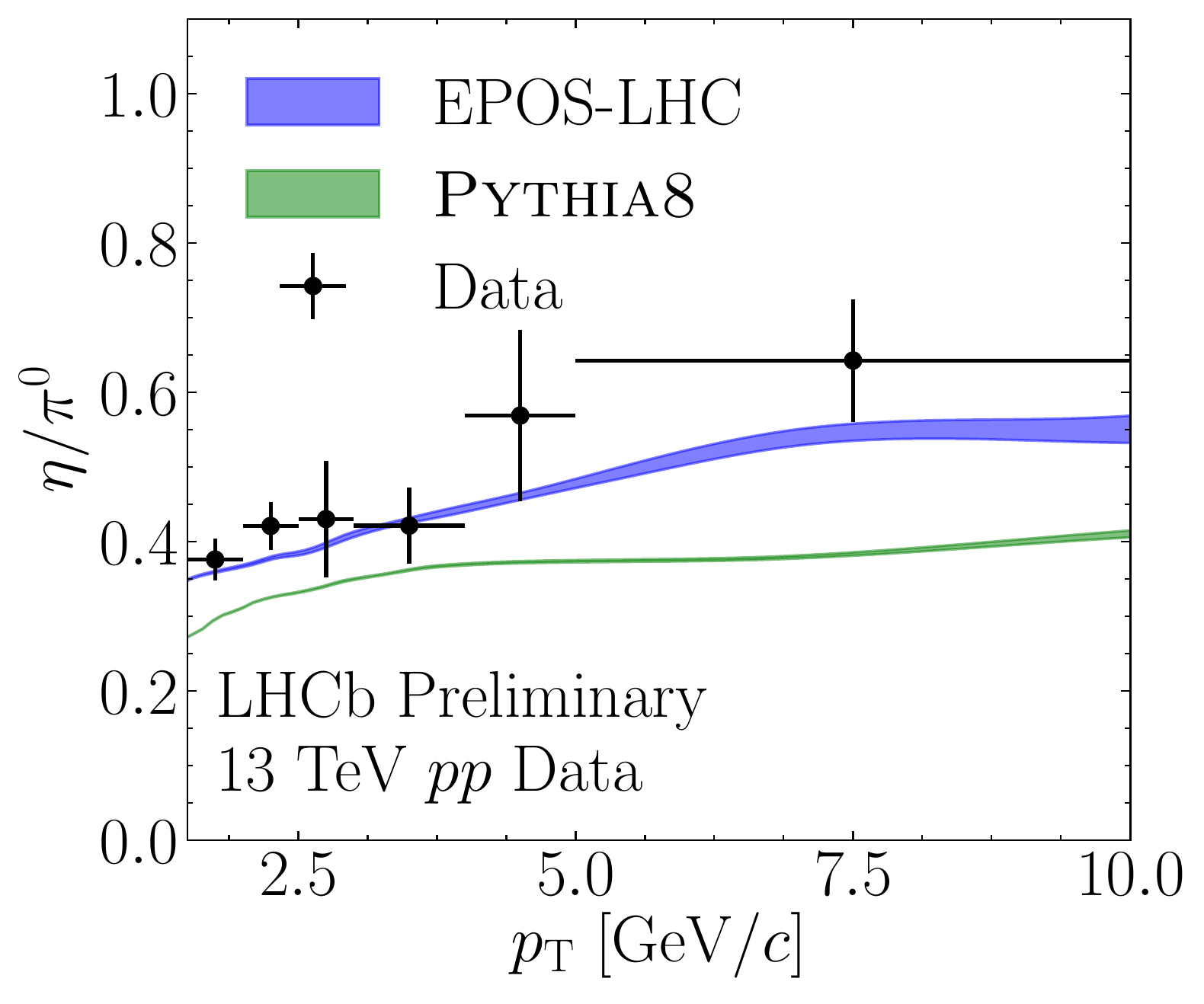}
    \includegraphics[width=0.40\textwidth]{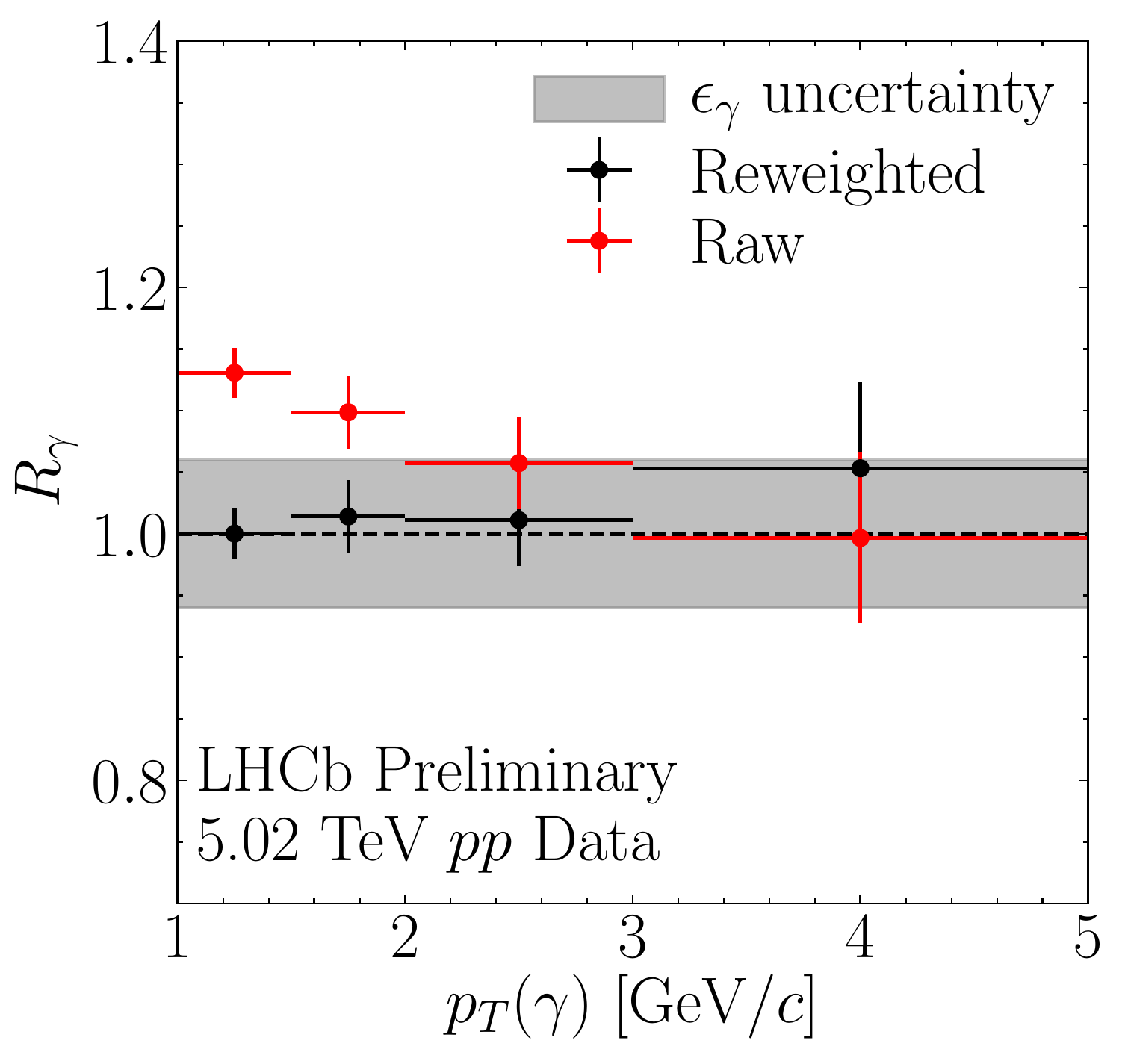}
        \caption{The $\eta$ over $\pi^0$ production ratio in $pp$ collisions as a function of transverse momentum and the auxiliary $R_{\gamma}$ ratio in $pp$ collisions without and with reweight corrections~\cite{perfplots}.}
    \label{fig:photon}
  \end{figure}

  LHCb pursues an effort using photons reconstructed in the $e^+e^-$ conversion channel. A target observable is the $R_{\gamma} = \frac{N_{incl, \gamma}^{Data}/N_{\gamma from \pi0}^{Data}}{N_{\gamma, decay}^{MC}/N_{\gamma,\pi 0}}$ double ratio extracting the direct photons as an excess over the decay photons inside the inclusive sample. The photons from $\pi^0$ are found by combining the conversion photons with a photon reconstructed in the calorimeter for a mass fit. For the measurements on isolated photons in $p$Pb, the $pp$ data sets without applying isolation criteria are regarded as control samples. Figure~\ref{fig:photon} shows the $\eta$ over $\pi^0$ production ratio that enters the definition of the $R$-ratio. The results in data are very close to the EPOS simulation ones. On the right hand side, the current double ratio $R$-measurement at 5~TeV using a simulation for 13~TeV $pp$ data is shown as a function of $\gamma$-$p_T$ with the present systematic uncertainty on the photon reconstruction efficiency, which is dominated by the extrinsic uncertainty of the branching fraction ratio $BR(B \to \chi_{c1}(\to J\psi +\gamma) K^+)/BR(B \to J/\psi K^+)$. This illustration demonstrates the understanding of different $pp$ data samples in terms of different kinematics and final state particle multiplicity at a level of the present systematic uncertainty without an excess being visible in this data set.    The measurements of $\gamma$-hadron correlations are also carried out as a complementary approach. Both measurements are unique chances to shed light in the low-$x$ frontier of QCD probed at the highest  energy hadron collider.
  
 \section{Coherent J/$\psi$ production in ultra-peripheral PbPb collisions}
  \label{sec:UPC}
\begin{figure}
    \includegraphics[width=0.6\textwidth]{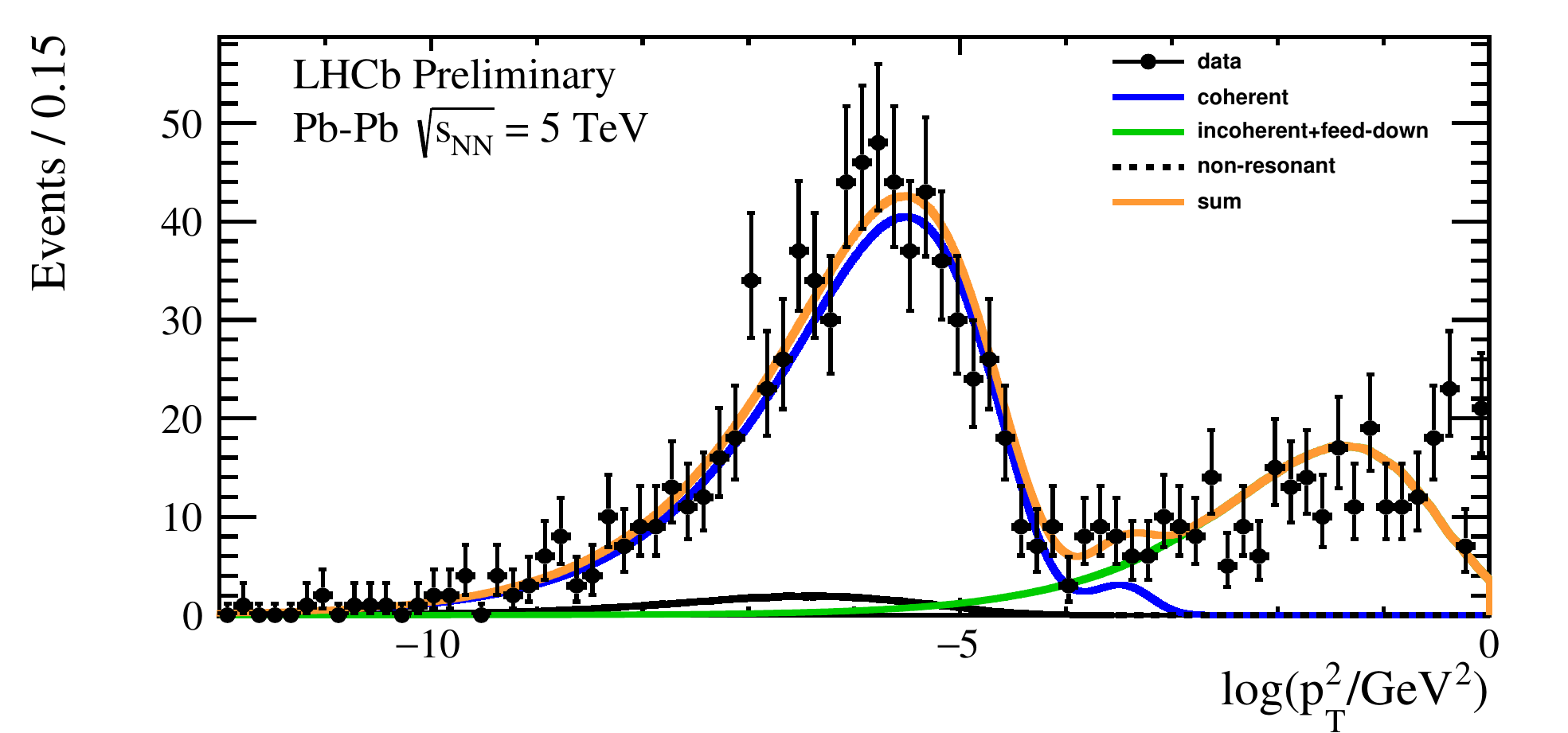}
    \includegraphics[width=0.39\textwidth]{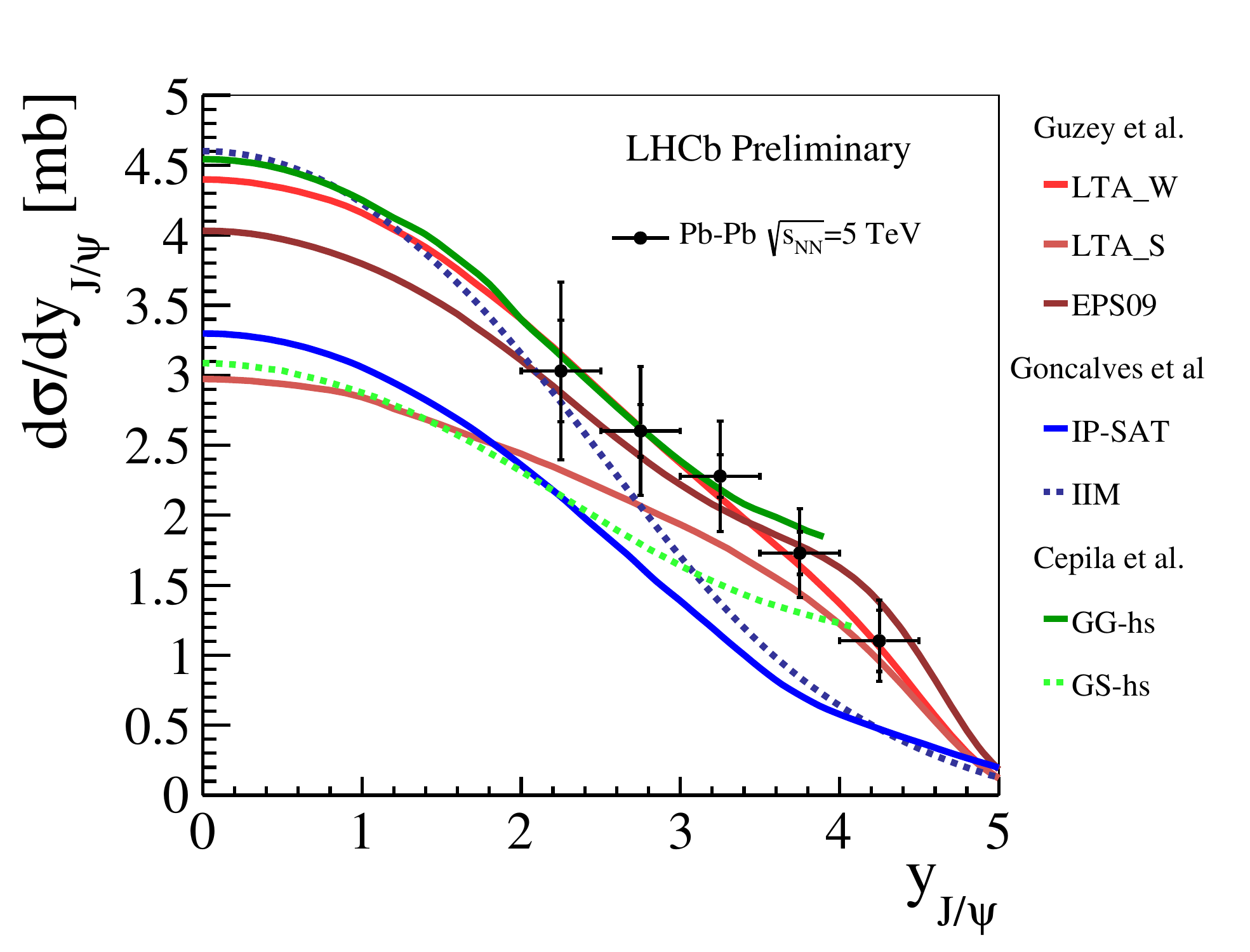}
    \caption{The $p_T^2$ distribution of dimuon candidates in the J/$\psi$ mass region as observed by LHCb. On the right hand side, the  extracted coherent J/$\psi$ production cross section is compared with model calculations~\cite{UPC}.}
    \label{fig:UPC}
  \end{figure}

  Ultra-peripheral collisions in PbPb collisions are an opportunity to probe the nucleus with the quasi-real photon cloud accompanying ultra-relativstic high-charge ions as $^{82+}$Pb. We report the first measurement of coherent J/$\psi \to \mu^+ \mu^-$ production by LHCb~\cite{UPC}. Figure~\ref{fig:UPC}  demonstrates the capability to separate the coherent from the incoherent production and the continuum. This measurement is sensitive to the gluonic content of the nucleus. On the right hand side, the results are compared with different families of model calculations. With the rapidity range covered by the experimental data points, it is possible to disfavour several model calculations in their present form: a detailed discussion can be found in~\cite{UPC}.  Measurements in ultra-peripheral PbPb collisions will strongly profit from the anticipated 10 times larger recorded integrated luminosity in LHCb for the 2018 run compared to the 2015 data set presented here.
  
  \section{Summary and Outlook}
  We have presented new results over a variety of different probes to characterise $pp$ and nuclear collisions at the LHC. First measurements of Bose-Einstein correlations  in $pp$ collisions have been presented and show qualitatively similar findings as previous measurements at midrapidity.  Results on open  heavy-flavour and quarkonium production  show strong suppressions at forward rapidity qualitatively compatible with strong shadowing within nPDFs, gluon depletion via saturation within the colour glass condensate framework or coherent energy loss. The $\Lambda_C$ production has been measured in $p$Pb collisions and its production ratio with respect to $D^0$ shows patterns compatible with a model tuned to the available production measurements in $pp$ collisions.  Steps towards measurements of photon production in $p$Pb collisions to probe the nuclear partons with a colour neutral final state in the quest for saturation are highlighted. In fixed-target collisions, first cross section measurements of $D^0$ production in $p$He collisions are shown. They do not show enhancements from data parameterisations and pQCD-based models that do not include intrinsic charm at the backward rapidity edge of the acceptance. Finally, first results from the 2015 PbPb data taking have been presented: coherent J/$\psi$ production in ultra-peripheral collisions to constrain the gluonic content of the nuclear wave function.

 These diverse measurements point to a large potential of LHCb to probe and characterise the initial state of heavy-ion collisions that is instrumental for the understanding of ion-ion collisions and that waits to be fully exploited in upcoming years. The planned first upgrade for LHC Run 3 improves the detector granularity that will be beneficial in particular for ion-ion collisions both in fixed-target collisions starting operations in 2021~\cite{LHCbCollaboration:2013bkh,LHCbCollaboration:2013iwy,LHCbCollaboration:2014tuj,LHCbCollaboration:2014vzo}.  The fixed-target collision programme will profit from a new internal gas target with 10-100 times higher useable instantaneous luminosity that is under consideration 
 by the LHCb collaboration at the same time scale.  The phase 2 upgrade planned to become operational around 2030,  for which an expression of interest has been submitted by the LHCb collaboration, bears the opportunity for an unprecented set-up in ion-ion collisions: a precision low-$p_T$ capable detector at forward rapidity using track momenta of minimal ionising particles and above.  These upcoming detector improvements for LHCb and the anticipated increased luminosities in the future ion-running at the LHC constitute a unique opportunity for heavy-ion physics  that should not be missed.

\textit{Acknowledgements}: The contact author acknowledges support from the 
European Research Council (ERC) through the project EXPLORINGMATTER,
founded by the ERC through a ERC-Consolidator-Grant. The contact author
thanks his collaborators for two years of excellent work.





\bibliographystyle{elsarticle-num}
\bibliography{<your-bib-database>}



\end{document}